\documentclass[12]{revtex4}

\usepackage{graphicx}

\begin{document}

\title{\textbf{Super-\v {C}erenkov Radiation: A new phenomenon useful for RICH
Detectors}}
\author{D. B. Ion$^{1,2)}$ and M. L. Ion$^{3)}$ \\
$^{1)}$TH Division, CERN, CH-1211 Geneva 23, Switzerland and\\
$^{2)}$NIPNE-HH , Bucharest P.O Box MG-6. Romania\\
$^{3)}$Bucharest University, Department of Atomic and Nuclear Physics,
Bucharest, Romania}
\date{}

\begin{abstract}
In this contribution the Super-\textit{\v {C}}erenkov radiation (SCR) as a
new phenomenon which includes in a more general and exact form the usual
\textit{\v {C}}erenkov effect is presented. The Super-\textit{\v {C}}erenkov
effect at \textit{\v {C}}erenkov threshold in the radiators of RICH
detectors is investigated. The results on the experimental test of the super-%
\textit{\v {C}}erenkov coherence conditions are presented. The S\textit{\v
{C}}R-predictions are verified experimentally with high accuracy $\chi
^{2}/n_{dof}=$1.47 by the data on the \textit{\v {C}}erenkov ring radii of
electron, muon, pion and kaon, all measured with RICH detector. Moreover, it
is shown that the Super-\textit{\v {C}}erenkov phenomenon can explain not
only subthreshold \v {C}R but also the observed secondary rings (or
anomalous \textit{\v {C}}erenkov radiation) observed at CERN SPS
accelerator. The influence of medium on the particle propagation properties
is also estimated and the refractive properties of electrons, muons, pions,
in the radiator C$_{4}$F$_{10}$Ar are obtained. So, we proved that the
refractive indices of the charged elementary particles in medium are also
very important for the RICH detectors, especially at low and intermediate
energies.
\end{abstract}

\maketitle

\textbf{Introduction. }The classical theory of the radiation
emitted by charged particles moving with superluminal velocities
were traced back to Thomson, Heaviside and Sommerfeld. In fact,
Heaviside [1] considered the \v {C}erenkov radiation in a
nondispersive medium. He considered this topic many times over the
next 20 years, deriving most of the formalism [2] of what is now
called \textit{\v {C}erenkov radiation} (\v {C}R). In 1904
Sommerfeld [3] considered radiation from a charge moving in a
vacuum at a velocity faster than light velocity ($v>c$) and close
approached the formulation of the theory of \v {C}R-effect. But
Sommerfeld did not applied his results for the particles motion in
a refractive (transparent) medium. Moreover, we must add that the
\textit{\v {C}}R have been observed before the \textit{\v
{C}}erenkov and Vavilov by Mallet and other authors but cannot be
regarded as the discovery of \textit{\v {C}}R since the essential
characteristic features of this radiation were not revealed and it
was not understood that the observed effect is clearly different
from luminescence. Consequently, the realizable case of radiation
from a charge moving with a constant velocity greater than the
phase velocity of light in a dielectric medium was discovered
independently in an experiment in 1934 by \v {C}erenkov [4]. So,
doing justice [2] to Heaviside and Sommerfeld, we must recall that
the classical theory of the \textit{\v {C}}R phenomenon in a
dispersive medium was first formulated by Frank and Tamm [4] in
1937. This theory explained all the main features of the radiation
observed by \v {C}erenkov [4]. The quantum theoretical approach to
the \v {C}R-problem was developed by many authors (see refs. in
[5]). The remarkable properties of the \v {C}erenkov radiation
find wide applications in practice especially in high energy
physics where it is extensively used in experiments for counting
and identifying relativistic particles in the fields of elementary
particles, nuclear physics and astrophysics. A short review of \v
{C}erenkov radiation and its use for particle identification with
threshold and differential counters is presented in Refs.[6]. In
the last decades the \v {C}erenkov radiation (\v {C}R) is the
subject of many studies related to extension to the nuclear media
[7]-[9] as well as to other coherent particle emission via \v
{C}erenkov-like mechanisms [10]. The generalized \v
{C}erenkov-like effects based on four fundamental interactions has
been investigated and classified recently in ref. [11]. In
particular, this classification includes the nuclear (mesonic,
$\gamma $, weak boson)-\v {C}erenkov-like radiations as well as
the high energy component of the coherent particle emission via
(baryonic, leptonic, fermionic)-\v {C}erenkov-like effects. In
1999, G.L.Gogiberidze, L.K. Gelovani and E.K. Sarkisyan performed
the first experimental test [12] of the pionic \v {C}erenkov-like
effect in Mg-Mg collisions at 4.3 GeV/c/nucleon and obtained a
good agreement with the position and width of the first pionic \v
{C}erenkov-like band predicted in ref. [11].

In essence, it was revealed (see ref. [4]) by Cerenkov, Tamm and
Frank that a charged particle moving in a transparent medium with
an refractive index $n_{\gamma },$ and having a speed $v_{x}$
greater than phase velocity of light $(v_{\gamma ph}=1/n_{\gamma
})$ will emit \v {C}erenkov radiation (\v {C}R) at an polar
emission angle $\theta _{C}$ relative to the direction
of motion given by the relation (we adopted the system of units $\hbar =c=1$%
)
\begin{equation}
\cos \theta _{\check{C}}=\frac{v_{\gamma ph}}{v_{x}}=\frac{1}{n_{\gamma
}v_{x}}\leq 1  \label{1}
\end{equation}

However, by recent experimental observations of the subthreshold [13] and
anomalous \v {C}erenkov radiations [14] it was clarified that some
fundamental aspects of the \v {C}R can be considered as being still open and
that more theoretical and experimental investigations on the \v {C}R are
needed. So, these new results stimulated new theoretical investigations
[15-16] (using the \v {C}R correct kinematics) leading to the discovery that
\v {C}R is in fact only a component (the low energy component) of a more
general phenomenon called by us the \textit{Super-\v {C}erenkov radiation}
(S\v {C}R) characterized by the \textit{Super-\v {C}erenkov coherence}
condition [16]
\begin{equation}
\cos \theta _{S\check{C}}=v_{xph}\cdot v_{\gamma ph}\leq 1  \label{2}
\end{equation}
where $v_{xph}$ and $v_{\gamma ph}$ are phase velocities of the
charged particle and photon in the medium, respectively. We must
underline that the discovery of the subthreshold \v {C}R was of
decisive importance for us to formulate the S\v {C}R-theory
[15-16] which includes the above mentioned new phenomena.
Moreover, we can see that the S\v {C}R coherence condition (2) is
obtained in a natural way from the energy-momentum conservation
law when the influence of medium on the propagation properties of
the charged particle is taken into account. Therefore, the problem
of the experimental test of \textit{Super-\v {C}erenkov coherence
condition (2)} is of great interest not only for the fundamental
physics but also for practical applications to the particle
detection. Recently such a test was performed [16] by using the
Ring Imaging Cherenkov (RICH) detectors.

Now, let us apply the theory of \textit{Super-\v {C}erenkov radiation} (S\v
{C}R) to predict some characteristic feature of the subthreshold \v
{C}erenkov-like effects in the RICH detectors.

\textbf{The Super-\v {C}erenkov Radiation (S\v {C}R) Theory. }Let start with
an electromagnetic decay

\begin{equation}
1\rightarrow \gamma +2  \label{3}
\end{equation}
in a (dielectric, nuclear or hadronic)-medium (we will work in the
system of units $\overline{h}=c=1),$ described in

\begin{center}
\noindent
\begin{minipage}{8.5cm}
\begin{center}
\includegraphics[width=6.5cm]{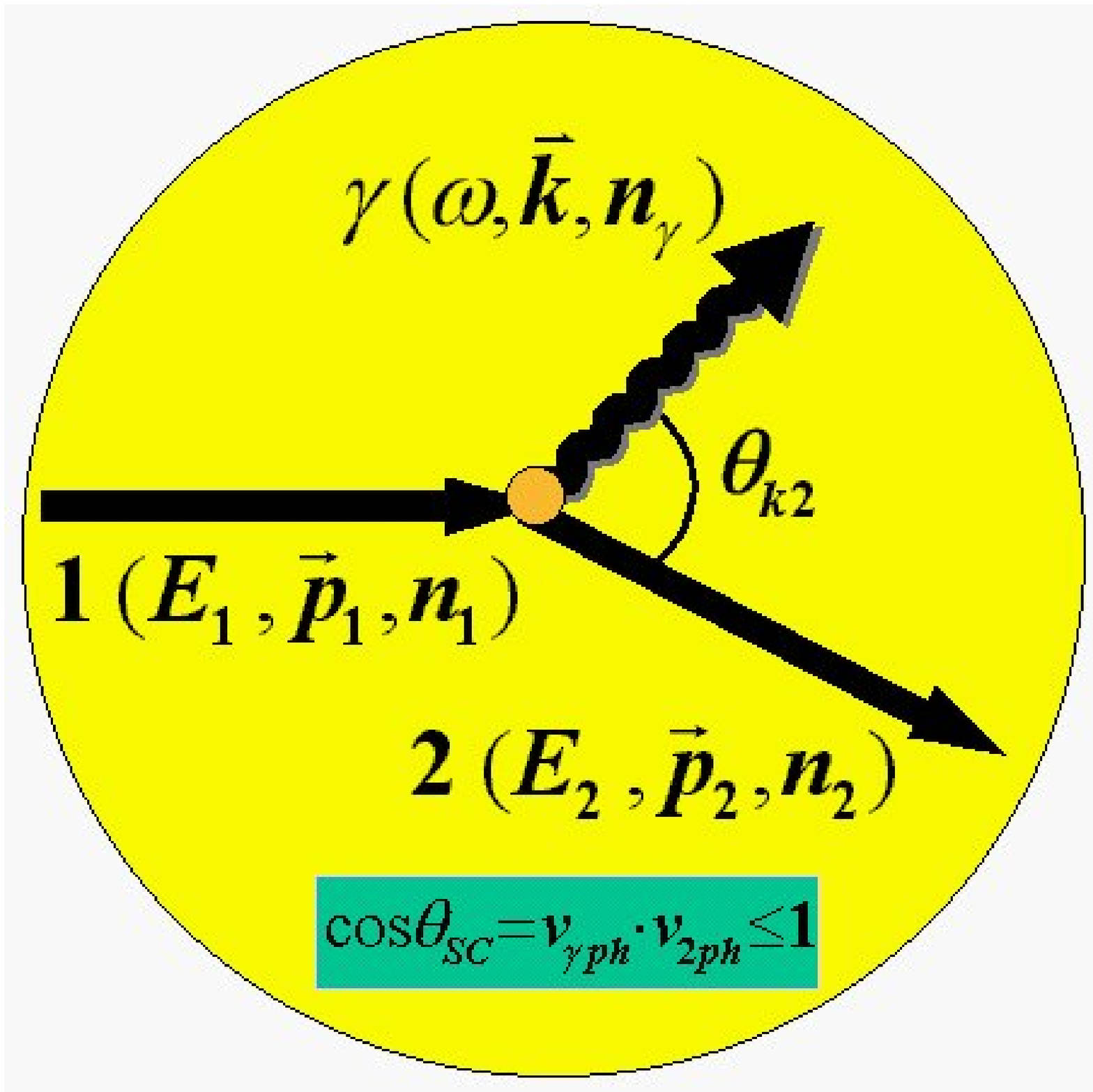}\\
\end{center}

{{\bf Figure 1:} \footnotesize A two-body decay in medium:
Super-Cerenkov coherence condition. }
\end{minipage}
\end{center}

Fig. 1, where a photon $\gamma $ [with energy $\omega ,$ momentum
$k=\omega {\rm Re}\;n_{\gamma }(\omega )$ and refractive index
$n_{\gamma }(\omega )$] is emitted in a (dielectric,
nuclear or hadronic) medium from an incident charged particle [with charge $%
Ze$, energy $E_{1},$ momentum $p_{1}={\rm Re}\;%
n_{1}(E_{1})(E_{1}^{2}-M^{2}]^{1/2}$ , rest mass M, the refractive index $%
n_{1}(E_{1})$] that itself goes over into a final particle $2$ [ with charge
Ze, energy $E_{2},$ momentum $p_{2}={\rm Re}\;%
n_{2}(E_{2})(E_{2}^{2}-M^{2}]^{1/2}$ , rest mass M, the refractive index $%
n_{2}(E_{2})$]. The refractive index $n_{x}(E_{x})$ of any particle x (with
the energy $E_{x},$ momentum $p_{x}$, rest mass M$_{x}$) in a medium
composed from the constituents c will be described in standard way by the
Foldy-Lax formula [17]
\begin{equation}
n_{x}^{2}(E_{x})=1+\frac{4\pi \rho }{E_{x}^{2}-M_{x}^{2}}\cdot C(E_{x})%
\overline{f}_{xc\rightarrow xc}(E_{x})  \label{4}
\end{equation}
where $\rho $ is the density of constituents c, $C(E_{x})$ is a coherence
factor, $\overline{f}_{xc\rightarrow xc}(E_{x})$ is the averaged forward
xc-scattering amplitude. $C(E_{x})=1$ when the medium constituents are
randomly distributed. The phase velocity $v_{phx}(E_{x})$ of any particle x
in medium is modified as follows:

\begin{equation}
v_{xph}(E_{x})=\frac{E_{x}}{p_{x}}=\frac{v_{xph}^{vac}(E_{x})}{{\rm
Re}\; n_{x}(E_{x})}  \label{5}
\end{equation}
Now, using the energy-momentum conservation

\begin{equation}
E_{1}=\omega +E_{2},\bigskip \smallskip \ \ \overrightarrow{p_{1}}=%
\overrightarrow{k}+\overrightarrow{p_{2}}  \label{6}
\end{equation}
we obtain
\begin{equation}
\cos \theta _{1\gamma }=v_{1ph}(E_{1})v_{\gamma ph}(\omega )+\frac{1}{2p_{1}k%
}\left[ -D_{1}+D_{2}-D_{\gamma }\right]  \label{7}
\end{equation}
\begin{equation}
\cos \theta _{12}=v_{1ph}(E_{1})v_{2ph}(E_{2})+\frac{1}{2p_{1}p_{2}}\left[
-D_{1}-D_{2}+D_{\gamma }\right]  \label{8}
\end{equation}
where $D_{x}$ , x$\equiv B_{1},B_{2},$ $\gamma ,$ are the mass shell
relations in medium for x-particle and are given by
\begin{equation}
D_{x}=E_{x}^{2}-p_{x}^{2}  \label{9}
\end{equation}
We note that the second terms from the right side of Egs (7)-(8)
can be considered as quantum correction to the first one. So,
the\textit{\ semiclassical angles} are given by: $\cos \theta
_{1\gamma }=v_{1ph}(E_{1})v_{\gamma ph}(\omega ),$ $\cos \theta
_{12}=v_{1ph}(E_{1})v_{2ph}(E_{2}),.$respectively.

(a) \textit{Semiclassical Theory of S\v {C}R. } A semiclassical theory of
S\v {C}R can be developed step by step in similar way with the classical
theory of \v {C}R [4] and here we present only some final results for the
case of a transparent nondispersive medium. So, if in the Maxwell equations
we introduce the duality relation v$_{x}$=v$_{xph}^{-1}(E_{x})$ in the
charge and current distribution, then it is easy to obtain that the
classical intensity of the \v {C}erenkov radiation can now be written in a
more exact form
\begin{equation}
\begin{tabular}{c}
$\frac{dN}{d\omega }(SCR)=Z^{2}\alpha L\sin ^{2}\theta _{1\gamma }$ \\
$=Z^{2}\alpha L\left[ 1-v_{1ph}^{2}(E_{1})v_{\gamma ph}^{2}(\omega )\right]
, $%
\end{tabular}
\label{10}
\end{equation}

for $p_{1}\approx p_{2},$ where $\frac{dN}{d\omega }(SCR)$ is the
number of photons emitted in the energy interval: $(\omega ,\omega
+d\omega ),$ and L is the length of path. Hence, at \v {C}erenkov
threshold we have

\begin{equation}
\frac{dN}{d\omega }(S\check{C}R)|_{CRthr}\approx Z^{2}\alpha
L\left[ 1-\left( \frac{1}{{\rm Re}\;n_{1}(E_{1})}\right)
^{2}\right] \label{11}
\end{equation}
Therefore, the existence of the subthreshold \v {C}erenkov
radiation can be obtained for ${\rm Re}\;n_{1}(E_{1})>1.$

(b)\textit{\ Quantum Theory of S\v {C}R }[16]. Now, we start with
a two-body spin ($1/2^{+}\rightarrow \gamma +1/2^{+}$) decay in a
(dielectric, nuclear, or hadronic) medium where the propagation
properties of all three
particles (see Fig. 1) are changed according to the eqs. (4)-(5). \textbf{\ }%
Next, for simplicity we consider that the same interaction Hamiltonian as in
the \v {C}R-theory with some modifications of the source fields in medium
can describe the coherent $\gamma -$emission in all S\v {C}R-sectors. Then,
we obtain that the intensity of the \textit{Super-\v {C}erenkov radiation}
for transparent (nonabsorbent) media can be written in the following general
form
\begin{equation}
\frac{d^{2}N_{S\check{C}}}{dtd\omega }=\frac{\alpha Z^{2}}{v_{1}}\frac{1}{%
|n_{B_{1}}|^{2}|n_{B_{2}}|^{2}|n_{\gamma }|^{2}}\frac{k}{\omega }\frac{dk}{%
d\omega }\,S\cdot \Theta \left( 1-\cos \theta _{SC}\right)  \label{12}
\end{equation}
where $N_{S\check{C}R}$ is the total number of the $S\check{C}R$-photons, $%
\Theta \left( 1-\cos \theta _{SC}\right) $ is Heaviside step function, while
the spin factor $S$ is given by
\begin{equation}
\begin{tabular}{c}
$S=\frac{(E_{1}+M)(E_{2}+M)}{4E_{1}E_{2}}\cdot $ \\
$\cdot \left[ \frac{p_{1}^{2}}{E_{1}+M}+\frac{p_{2}^{2}}{E_{2}+M}+2\frac{(%
\vec{e}_{k}.\vec{p}_{1})(\vec{e}_{k}.\vec{p}_{2})-(\vec{e}_{k}\times \vec{p}%
_{1})(\vec{e}_{k}\times \vec{p}_{2})}{(E_{1}+M)(E_{2}+M)}\right] $%
\end{tabular}
\label{13}
\end{equation}
where the vector $\vec{e}_{k}$ is the photon polarization vector for a given
photon momentum $\vec{k}$. As in the usual case of \v {C}R-theory, for a
given vector $\vec{k}$ we choose two orthogonal photon spin polarization
directions, corresponding to a polarization vector $\vec{e}_{k}$
perpendicular and parallel to the decay plane Q given by the vectors $\vec{p}%
_{1}$ and $\vec{k}$, respectively. Then, from eq. (13) we get the following
expressions of the spin factors $S^{\bot }$ and $S^{||}$ :
\begin{equation}
S^{\bot }=\frac{(E_{1}+M)(E_{2}+M)}{4E_{1}E_{2}}\cdot \left[ \frac{\vec{p}%
_{1}}{E_{1}+M}-\frac{\vec{p}_{2}}{E_{2}+M}\right] ^{2}  \label{14}
\end{equation}
\begin{equation}
S^{||}=v_{1}\mathrm{Re}\,n_{1}\;v_{2}\mathrm{Re}\,n_{2}\;\sin \theta
_{1\gamma }\;\sin \theta _{2\gamma }  \label{15}
\end{equation}
where v$_{i},$ i=1,2 are the corresponding particle velocities in vacuum..

\begin{center}
\noindent
\begin{minipage}{14.5cm}
\begin{center}
\includegraphics[width=14.5cm]{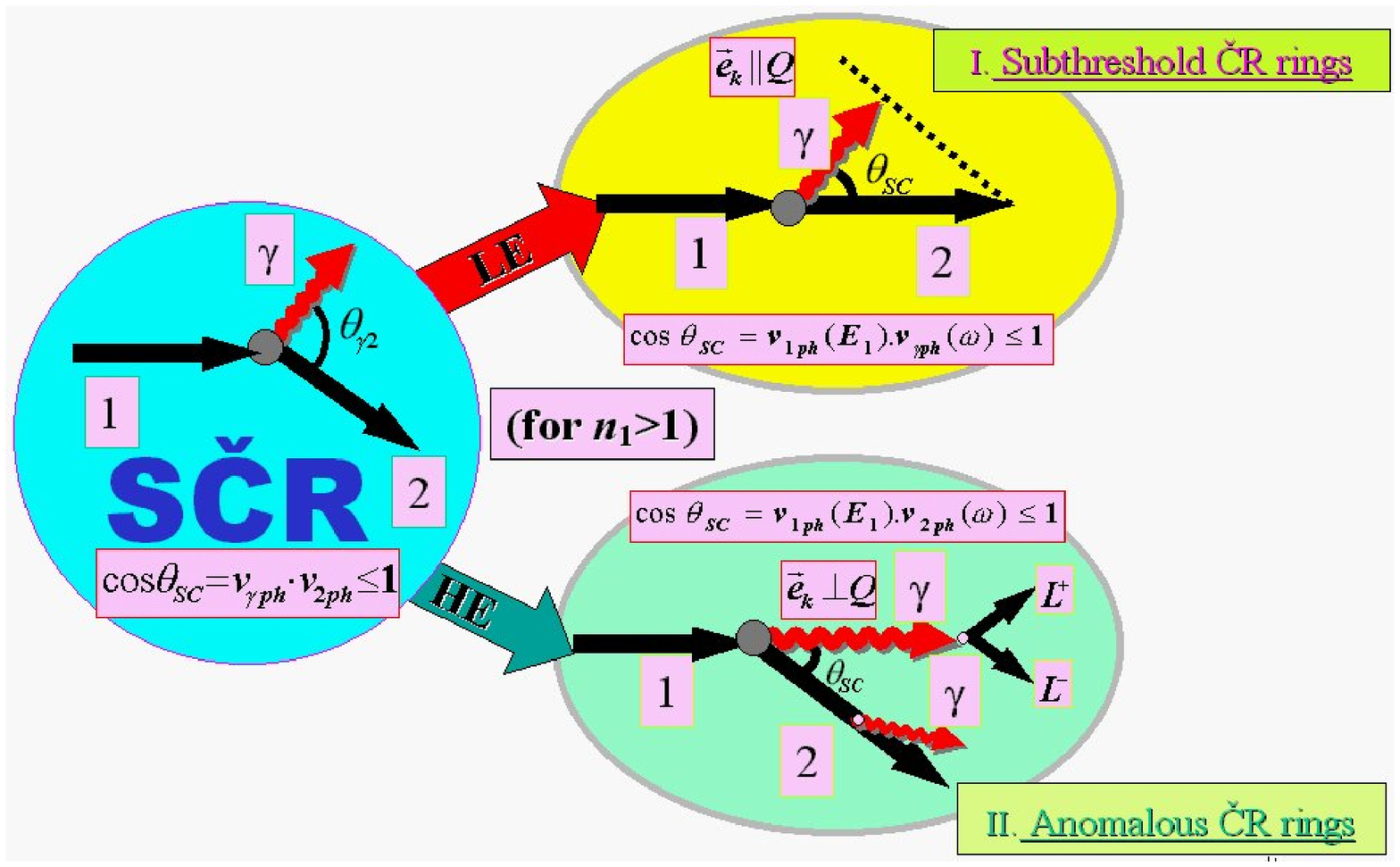}\\
\end{center}

{{\bf Figure 2:} \footnotesize Schematic description of the
Super-\v {C}erenkov radiation. }
\end{minipage}
\end{center}

Now, one can see that Heaviside function $\Theta \left( 1-\cos \theta
_{SC}\right) $ is 1 at least in two physical $\gamma $-energy regions
defined by the inequality (see Fig. 2): $\cos \theta _{2\gamma }\approx
v_{\gamma ph}(\omega )v_{2ph}(E_{2})\leq 1$. Hence, the low $\gamma -$energy
sector is that where $\theta _{SC}\equiv \theta _{2\gamma }\approx \theta
_{1\gamma }$, while the high $\gamma -$energy sector is that where $\theta
_{SC}\equiv \theta _{2\gamma }\approx \theta _{12}$. Therefore, the limiting
low $\gamma -$energy S\v {C}R-sector can be identified as extended $\gamma -$%
\v {C}erenkov domain where the condition: $v_{\gamma ph}(\omega
)v_{1ph}(E_{2})\leq 1$ is fulfilled, while limiting high $\gamma -$energy
S\v {C}R-sector can be identified as extended source-\v {C}erenkov-like
domain, in the sense that the charged particle spontaneously decays into a
high $\gamma -$energy photon fulfilling a generalized \v {C}erenkov-like
relation: $v_{1ph}(\omega )v_{2ph}(E_{2})\leq 1$.

\begin{center}
\noindent
\begin{minipage}{14.5cm}
\begin{center}
\includegraphics[width=14.5cm]{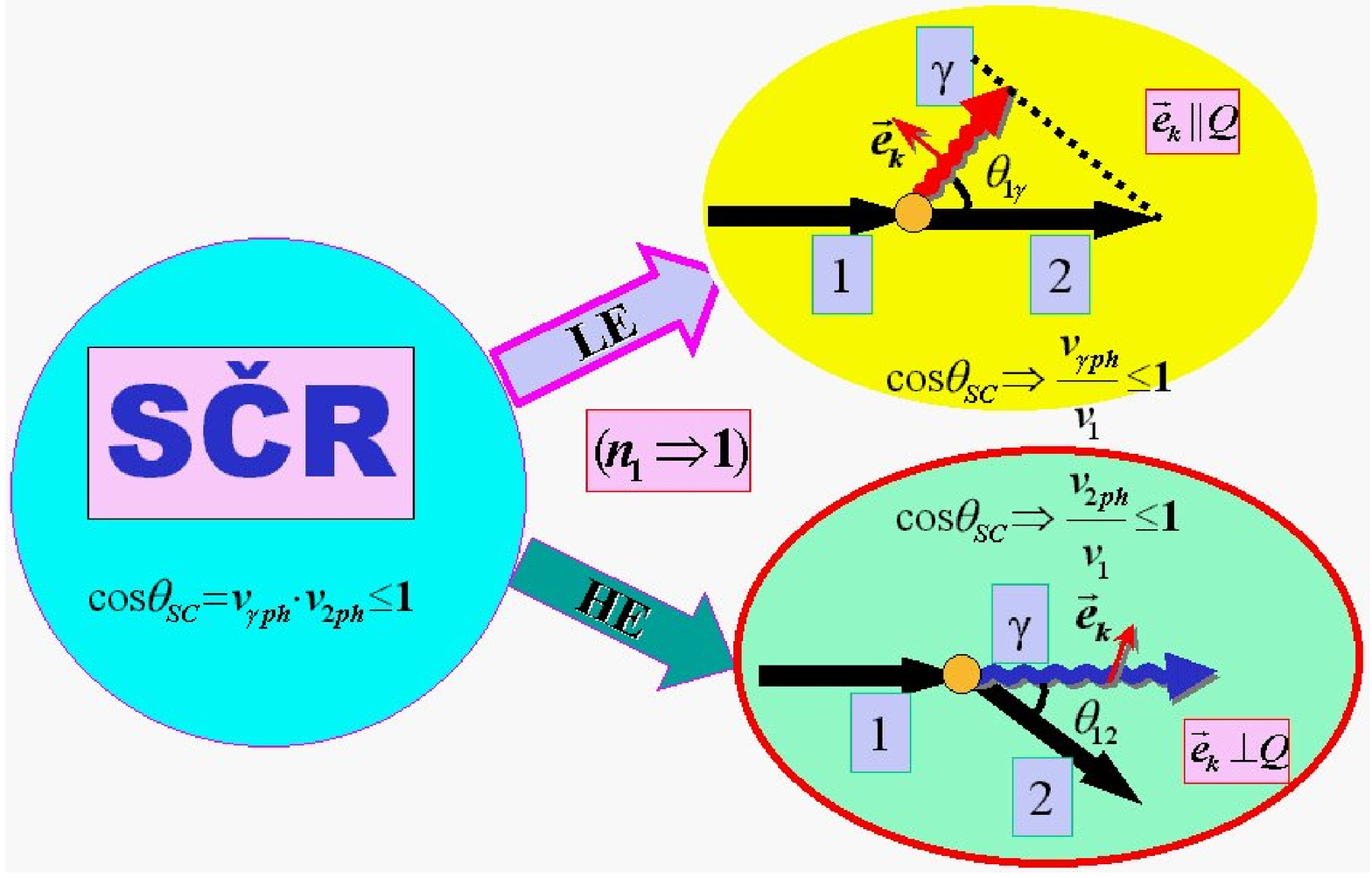}\\
\end{center}

{{\bf Figure 3:} \footnotesize Unified description of two \v
{C}erenkov-like limits of the Super-\v {C}erenkov radiation when
$n_{1}(E_1)=1$. }
\end{minipage}
\end{center}

Of course, in the limit $n%
_{1}(E_{1})\rightarrow 1$ these physical regions are going (see Fig. 3) in
the corresponding \v {C}R-like domains where the respective conditions: $%
v_{\gamma ph(\omega )\leq }v_{1}$ and $v_{2ph(E_{2})\leq }v_{1}$,
are satisfied. In conclusion the Super-\v {C}erenkov effect cannot
be confused with the usual \v {C}erenkov radiation since the S\v
{C}R can be considered as a continuous two-body decay in medium
which includes not only two generalized \v {C}erenkov-like
phenomena but also their interference effects. The relations
(12)-(15) includes in a general and unified way all the main
predictions of the Super-\v {C}erenkov radiation from which the
results from Table 1 are obtained as two particular limiting
cases.

\begin{center}
\noindent {\bf Table 1:} The  S\v CR  main predictions.
\nopagebreak

{\footnotesize \noindent
\begin{tabular}{|p{2mm}|p{1.3cm}|p{3.0cm}|p{2.9cm}|} \cline{1-4}
  & Name & \parbox[c]{2.6cm}{\centering (S\v CR) Low $\gamma$-energy sector \\(see LE in figs. 2-3)} &
 \parbox[c]{2.4cm}{\centering (S\v CR) High $\gamma$-energy sector} (see HE in figs. 2-3)\\ \cline{1-4}
1 &  \parbox[c]{1.4cm}{coherence relation} & $v_{\gamma ph}\cdot
v_{xph}\leq 1$ & $v_{\gamma ph}\cdot v_{xph}\leq 1$ \\ \cline{1-4}
2 &  \parbox[c]{1.4cm}{coherence angle} &
\parbox[c]{2.7cm}{\centering $\cos\theta _{SC}=v_{\gamma ph}\cdot
v_{1ph}$ since $\theta_{SC}\equiv \theta_{2\gamma}\approx
\theta_{1\gamma}$} &
 \parbox[c]{2.7cm}{\centering $\cos\theta _{SC}=v_{1 ph}\cdot v_{2ph}$
since $\theta_{SC}\equiv \theta_{2\gamma}\approx \theta_{12}$} \\
\cline{1-4} 3 &  \parbox[c]{1.4cm}{threshold velocity} &
\parbox[c]{2.7cm}{\centering $v_{xthr}(SC)=\frac{1}{n_1n_\gamma}$}
&
 \parbox[c]{2.7cm}{\centering $v_{xthr}(SC)=\frac{1}{n_1n_2}$} \\ \cline{1-4}
4 &  \parbox[c]{1.4cm}{maximum emission angle} &
\parbox[c]{2.7cm}{\centering
$\theta_{SC}^{max}=\arccos\frac{1}{n_1n_\gamma}$} &
 \parbox[c]{2.7cm}{\centering $\theta_{SC}^{max}=\arccos\frac{1}{n_1n_2}$} \\ \cline{1-4}
5 &  \parbox[c]{1.4cm}{\vspace{1mm} spectrum \\} &
\parbox[c]{2.9cm}{\centering $\frac{dN_{\gamma}}{d\omega }=\alpha
LZ^2 \sin^2\theta_{SC}$} &
 \parbox[c]{2.7cm}{\centering $\frac{dN_{\gamma}}{d\omega }=LZ^2/2\;^*)$} \\ \cline{1-4}
6 &  \parbox[c]{1.4cm}{polari-\\zation} & 100\% ($\vec{e}||Q$) &
100\% ($\vec{e}\bot Q$)  $^{**})$\\ \cline{1-4}
\end{tabular}}
\end{center}

The results of quantum theory presented here are complete for the
Super-\v
{C}erenkov radiation (S\v {C}R) produced by any spin-1/2 particle [such as ($%
e^{\pm }$, $\mu ^{\pm }$, $\tau ^{\pm }$)-leptons, ($p$, $\Sigma ^{\pm }$, $%
\Xi ^{-}$, $\Omega ^{-}$)-baryons etc.] moving in a (dielectric, nuclear or
hadronic)-medium, with its phase velocity satisfying the Super-\v {C}erenkov
coherence condition (2).

\textbf{Experimental tests and predictions for Super-\v {C}erenkov radiation.%
} The problem of the experimental test of \textit{Super-\v {C}erenkov
coherence condition (2)} is of great interest not only for the fundamental
physics but also for practical applications to the particle detection. Such
a test can be performed by using a Ring Imaging \v {C}erenkov (RICH)
detectors (see ref. [18]).

\begin{center}
\noindent
\begin{minipage}{8.5cm}
\includegraphics[width=8cm]{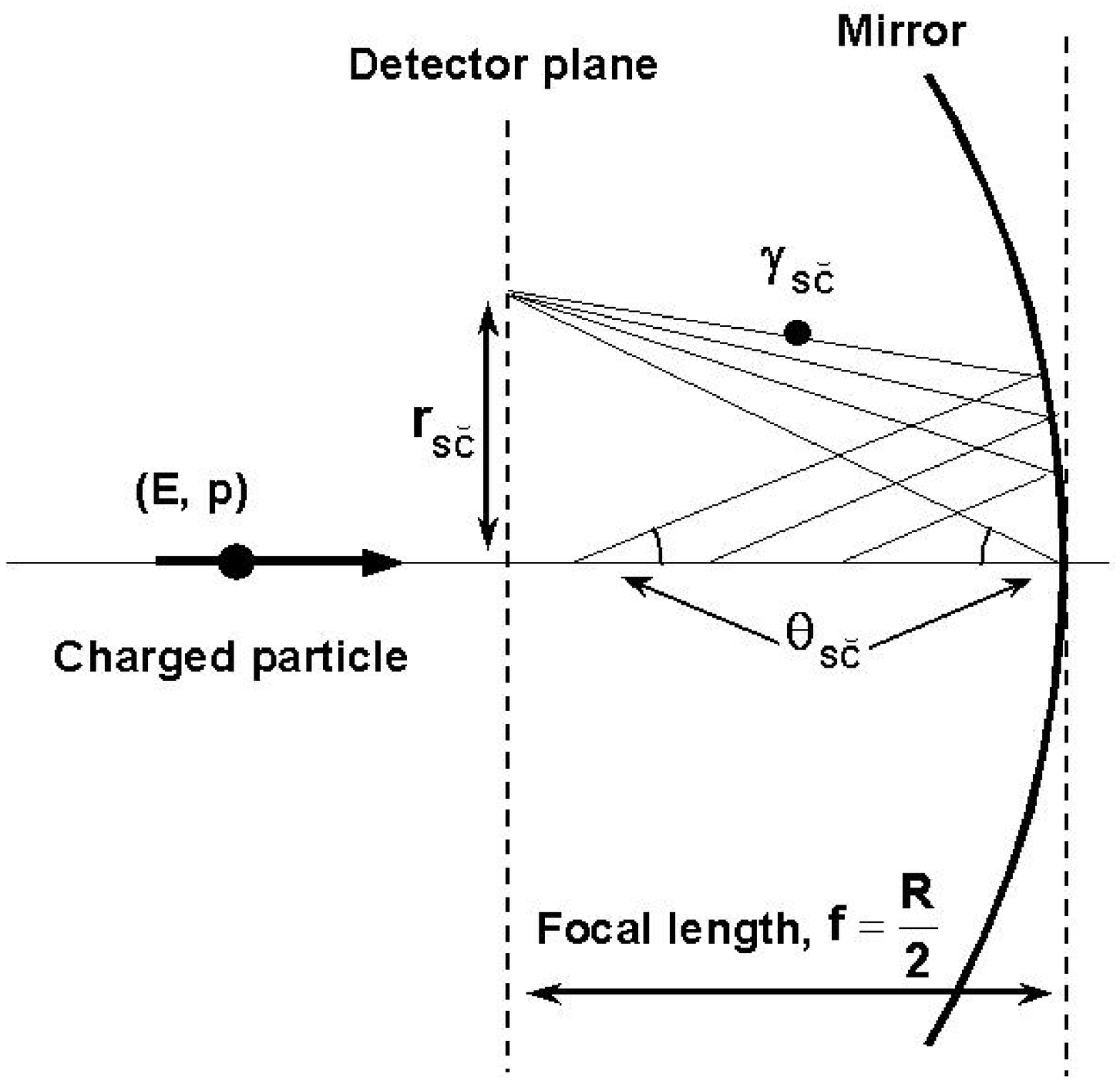}\\

{{\bf Figure 4:} \footnotesize Schematic description of a RICH
detector. }
\end{minipage}
\end{center}

In RICH detectors, particles pass through a
radiator, and a spherical mirror focuses all photons emitted at $\theta _{S%
\check{C}}$ along the particle trajectory at the same radius $%
r_{SC}=(R/2)\tan \theta _{SC}$ on the focal plane. Photon sensitive
detectors placed at the focal plane detect the resulting ring images in the
RICH detector. So, RICH-counters are used for identifying and tracking
charged particles. \v {C}erenkov rings formed on a focal surface of the RICH
provide information about the velocity and the direction of a charged
particle passing the radiator. The particle's velocities are related to the
Cerenkov angle $\theta _{C}$ or to the Super- \v {C}erenkov $\theta _{SC}$
by the relation (1) (or (2), respectively). Hence, these angles are
determined by measuring the radii of the rings detected with the RICH. In
ref. [18] a $C_{4}F_{10}Ar(75:25)$ filled RICH-counter read out was used for
measurement of the \v {C}erenkov ring radii. Fig. 5a shows the experimental
values of the ring radii of electrons, muons, pions and kaons measured in
the active area of this RICH-detector. The saturated light produced from
electrons was a decisive fact to take an index of refraction $n_{\gamma
}=1.00113$ for the radiator material. The absolute values for excitation
curves of electron, muon, pion and kaon, shown by dashed curves in Fig. 5a,
was obtained by using this value of refractive index in formula: $%
r_{C}(p)=(R/2)\tan \theta _{C}(p)$. The solid curves show the individual
best fit of the experimental ring radii with eq.. $r_{SC}(p)=(R/2)\tan
\theta _{SC}(p)$(see Table 1).

\begin{center}
\noindent {\bf Table 2:}  The best fit parameters of experimental
ring radii with the Super-\v Cerenkov prediction. \noindent
\begin{tabular}{|c|c|r|r|}
\hline Particle & \parbox[c]{1.4cm}{Number of exp.
data} & $10^3\cdot a^2$ (GeV/c)&$\chi^2/n_{dof}$\\
\hline
$e$   & 6 & -0.081$\pm$0.101 & 0.468 \\
$\mu$ & 4 & 1.449$\pm$0.098 & 3.039 \\
$\pi$ & 7 & 2.593$\pm$0.167 & 0.234 \\
$K$   & 1 & 21.140$\pm$2.604 & $<10^{-14}$ \\ \hline All
data & 18 & \parbox[c]{2.4cm}{$(a/m)^2=$ $0.1211\pm 0.0053$} & 1.47 \\
\hline
\end{tabular}
\end{center}

For the particle refractive index we used the parametrization
\begin{equation}
n_{x}^{2}(p)=1+a^{2}/p^{2},\;\;v_{x}=p/\sqrt{p^{2}+m^{2}}  \label{16}
\end{equation}
where p is the particle momentum in the vacuum.. In the paper [16]
wee fitted all the 18 experimental data on the ring radii from
ref.[18] with our Super-\v {C}erenkov prediction formula (see fig.
4)
\begin{equation}
r_{SC}(p/m)=\frac{R}{2}\tan \theta _{SC}  \label{17}
\end{equation}
and we obtained the following consistent result (see Fig. 5b). The
best fit parameters are as follow: $(a/m)^{2}=0.12109\pm 0.00528$
and $\chi ^{2}/n_{dof}=1.47$, where $n_{dof}=16$ is the number of
degree of freedom (dof). The $r_{SC}(p/m)$ scaling function [16]
together with all experimental data on the ring radii of the
electron, muon, pion and kaon, are plotted as a function of the
scaling variable (p/m) in Figs. 5b. Now combining eqs.(10) and
(16) we obtain

\[
\frac{dN}{d\omega }(S\check{C}R)=Z^{2}\alpha L\left[
1-\frac{1}{n_{\gamma }^{2}}\cdot
\frac{p^{2}+m^{2}}{p^{2}+a^{2}}\right]
\]

Therefore, at \v {C}R threshold we get:
\begin{equation}
\frac{dN}{d\omega }(S\check{C}R)|_{CRthr}\approx Z^{2}\alpha L\left[ \frac{%
\left[ \frac{a_{x}}{m_{x}}\right] ^{2}(n_{\gamma }^{2}-1)}{1+\left[ \frac{%
a_{x}}{m_{x}}\right] ^{2}(n_{\gamma }^{2}-1)}\right]  \label{18}
\end{equation}

or
\begin{eqnarray}
\frac{\frac{dN}{d\omega }(S\check{C}R)|_{CRthr}}{\frac{dN}{d\omega }(S\check{%
C}R)|_{v_{1}=1}}=\left[ \frac{a_{x}}{m_{x}}\right] ^{2}.\frac{n_{\gamma }^{2}%
}{1+\left[ \frac{a_{x}}{m_{x}}\right] ^{2}(n_{\gamma }^{2}-1)}
 =0.1301\pm 0.088   \label{19}
\end{eqnarray}
since

\begin{equation}
\begin{tabular}{c}
$p_{thr}(\check{C})/m=\frac{1}{(n_{\gamma }^{2}-1)^{1/2}}=21.029$ \\
$p_{thr}(S\check{C})/m$=$\frac{1}{(n_{\gamma }^{2}-1)^{1/2}}\left[
1-\left[
\frac{a_{x}}{m_{x}}\right] ^{2}n_{\gamma }^{2}\right] ^{1/2}=18.477$%
\end{tabular}
\label{20}
\end{equation}

Moreover, we can estimate the $\theta _{SC}$ and the S\v {C}-ring radius r$%
_{SC}$ at \v {C}erenkov thresholds and we obtain:

\begin{eqnarray}
\theta _{SC}(v_{Cthr})=\arctan \left[ \left[
\frac{a_{x}}{m_{x}}\right] (n_{\gamma }^{2}-1)^{1/2}\right]
=0.948\pm 0.021(\deg )  \label{21}
\end{eqnarray}

\begin{eqnarray}
r_{SC}(\check{C}R-thr)=\frac{R}{2}\left[ \left[
\frac{a_{x}}{m_{x}}\right] ^{2}(n_{\gamma }^{2}-1)\right] ^{1/2}
 =(1.466\pm 0.055)\,\,cm  \label{22}
\end{eqnarray}

\begin{center}
\noindent
\begin{minipage}{14.5cm}
\begin{center}
\includegraphics[width=8.cm]{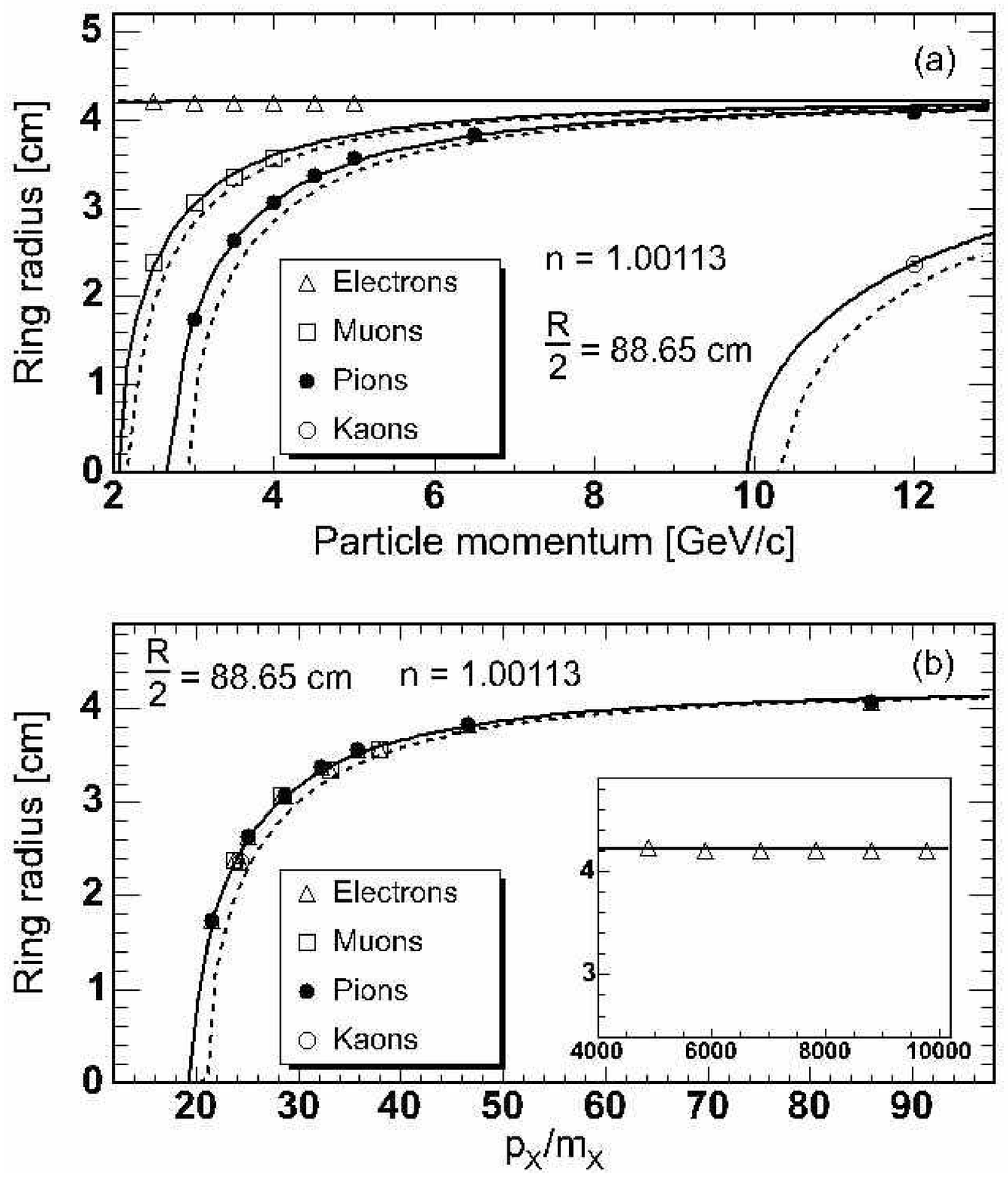}\\
\end{center}
{{\bf Figure 5:} \footnotesize Experimental test of SCR- coherence
condition (2) by using (17) and \v {C}erenkov ring radii obtained
by Debbe et al [18] with a RICH detector. a) SCR-prediction (solid
curve) and CR-prediction (dashed curve). b) The SCR-scaling (solid
curve) and CR-scaling (dashed curve) compared with ring radii (see
text). }
\end{minipage}
\end{center}

\begin{center}
\noindent
\begin{minipage}{14.5cm}
\begin{center}
\includegraphics[width=8.cm]{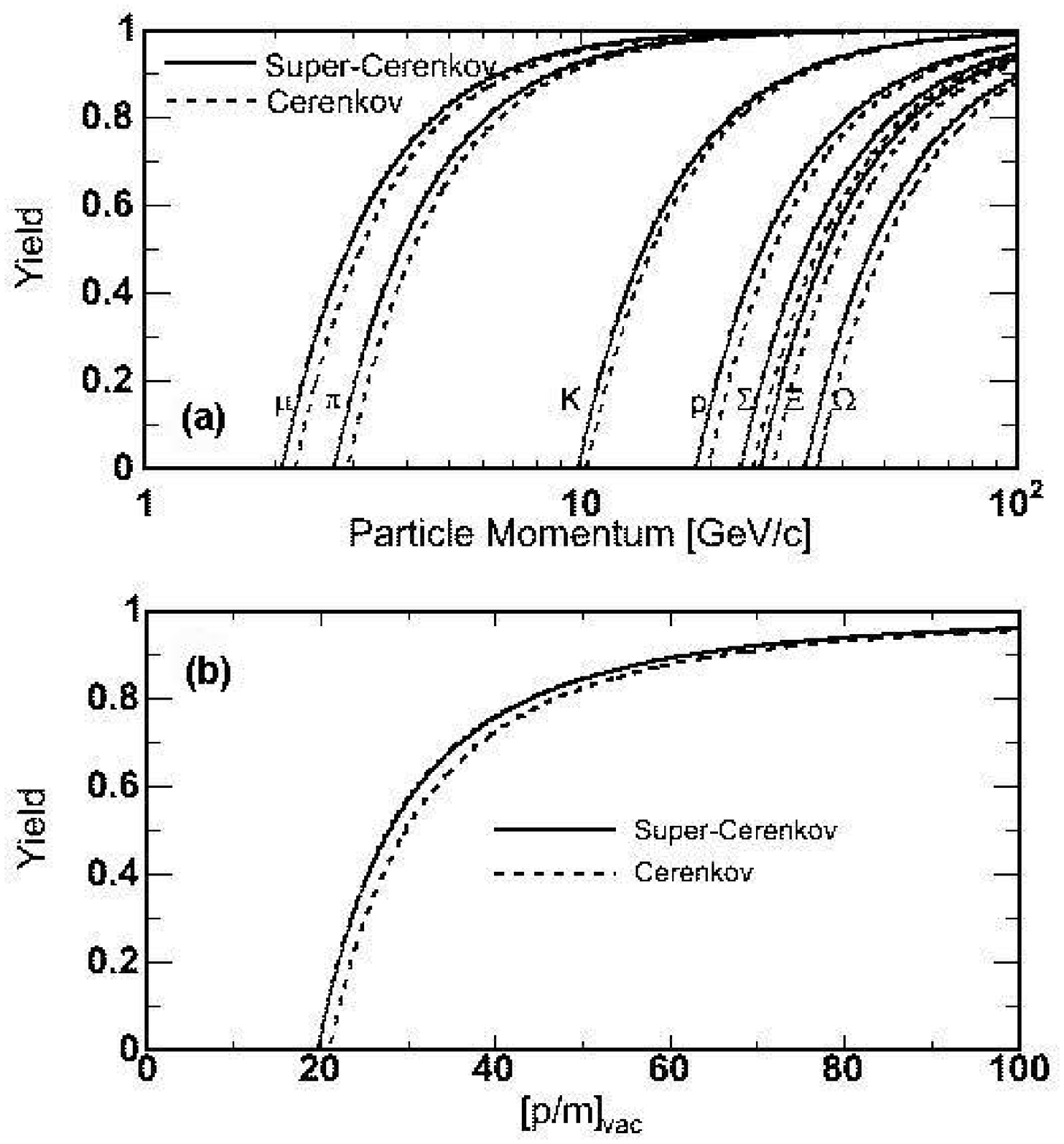}\\
\end{center}

{{\bf Figure 6:} \footnotesize The relative Yields (a) and their
scaling property (b) for the SCR-effect predicted by eqs. (10),
(16) and the fitted constant $(a/m)^2=0.12109$. }
\end{minipage}
\end{center}

In Figs. 6 we present the results for S\v {C}R-Yields defined as Y$_{SCR}$%
(p)=N$_{SCR}(p)/$N$_{SCR}$($\infty $). Also the difference $D_{Y}$(p)$=$Y$%
_{SCR}$(p)$-$Y$_{CR}$(p) is calculated and given in Fig.7. What is
interesting is that D$_{T}$(p) is maximum at \v {C}erenkov
threshold (see eq. (20)).

\textbf{Anomalous \v {C}R: }The anomalous \v {C}erenkov ring
observed recently by Vodopianov et al.[14] at SPS accelerator at
CERN Pb-beam can be considered also as experimental signature of
the HE-Super-\v {C}erenkov component (see Fig. 4) since both Pb
and high-energy $\gamma $ after spontaneous emission of a photon
can produce secondary anomalous \v {C}R-rings. One of the
characteristic feature of the predicted by S\v {C}R-effect (see
Fig. 4) is that the secondary anomalous \v {C}R-rings must have an
constant inclination angle $\alpha $ relative to beam direction
given by the relation

\begin{equation}
\cos \alpha =\cos \theta _{12}=v_{1ph}(E_{1})\cdot v_{2ph}(E_{2})
\label{23}
\end{equation}
This first important condition is verified experimentally with high accuracy
by four from seven anomalous \v {C}R-rings observed in ref. [14]. Therefore.
these \v {C}R-anomalous rings can be interpreted as being produced by the
photons emitted by secondary Pb (see again Fig.4). Then, it is easy to show
that the \v {C}erenkov angles cos$\theta ^{^{\prime }}$ must be given by a
relation of form:

\begin{equation}
\cos \theta _{2\gamma _{^{\prime }}}^{^{\prime }}=v_{\gamma ph}(\omega
)\cdot v_{2ph}(E_{2})=\frac{1}{n_{\gamma }}\cdot \frac{1}{n_{2}v_{2}}
\label{24}
\end{equation}
Therefore, the velocities higher than unity inferred by Vodopianov
et al. [14] from their anomalous \v {C}R-rings, must be divided by
the refractive index $n_{2}>1$ of the secondary Pb. Consequently,
their anomalous \v {C}R-rings [14] cannot be interpreted as being
produced by tachyons.

\begin{center}
\noindent
\begin{minipage}{8.5cm}
\includegraphics[width=8.5cm]{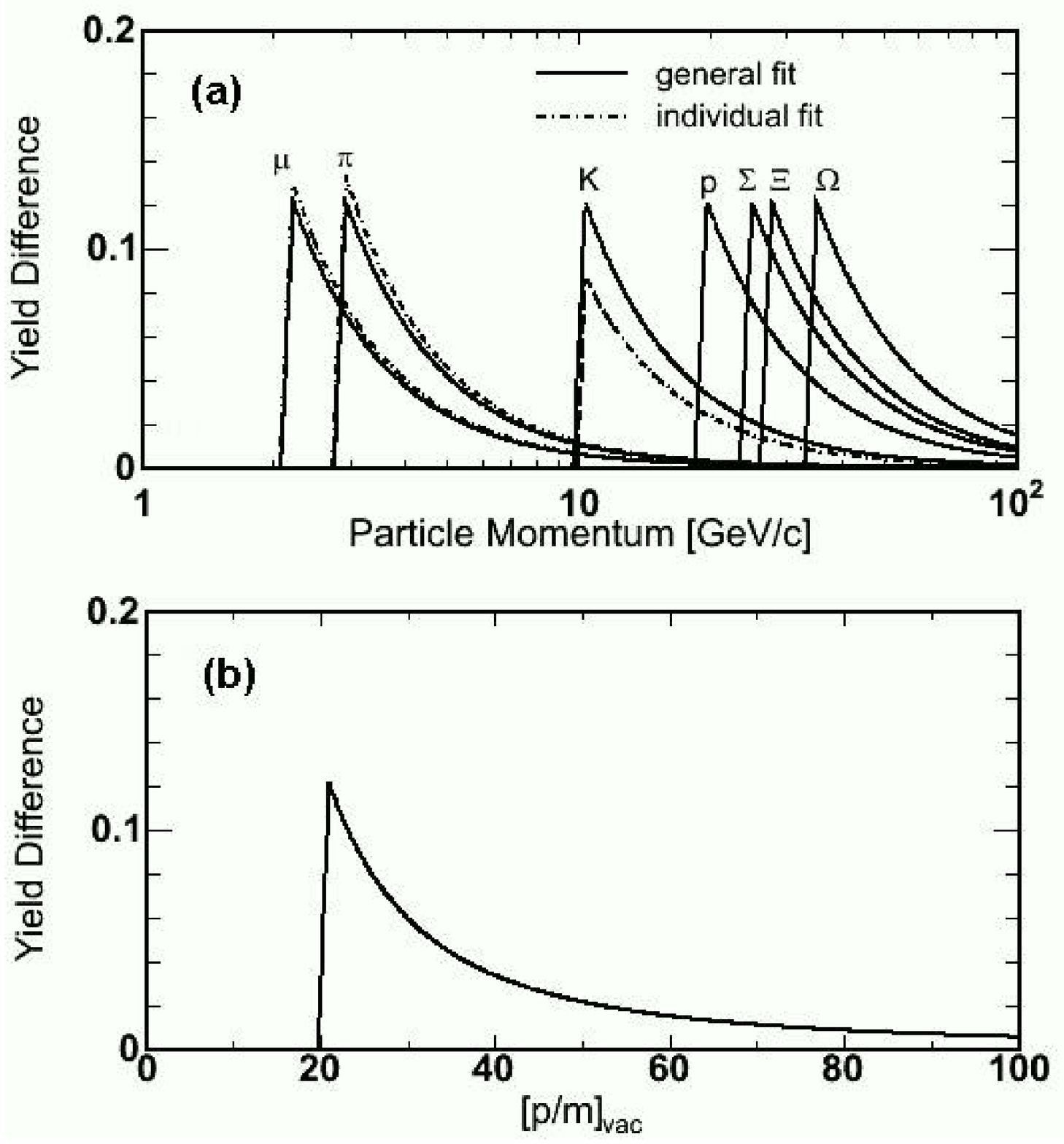}\\

{{\bf Figure 7:} \footnotesize The relative (a) Yields differences
$Y(SCR)-Y(CR)$ and (b) scaling of Yields differences. }
\end{minipage}
\end{center}

\textbf{Summary and Conclusions. }In this contributed paper the
Super-\v {C}erenkov effect as a new dual coherent particle
production mechanism is presented. The main results and
conclusions can be summarized as follows:

(i)  The Super-\v {C}erenkov phenomenon can be considered as a
continuous two-body decay in medium which is possible only in two
distinct limiting physical regions where the Super-\v {C}erenkov
coherence condition (2) can be fulfilled. One of them is at very
low $\gamma $-energies (LE)\ where
\begin{equation}
v_{1ph}^{-1}(E_{i})\geq v_{\gamma ph}(\omega ),\text{(extended }\gamma -%
\check{C}R\text{ region)}  \label{25}
\end{equation}
(see Figs. 2-3 and predictions in Table 1), and, a second region at very
high $\gamma $-energies (HE) where
\begin{equation}
v_{2ph}^{-1}(E_{2})\geq v_{1ph}(E_{1}),\text{ (extended }2-\text{\v
{C}erenkov-like region)}  \label{26}
\end{equation}
(see Figs. 2-3 and predictions in Table 1) where E$_{xi}$ and E$_{xf}$ are
the particle energy before and after $\gamma $-emission.

(ii) The experimental test of the S\v {C}R-coherence condition:
$v_{\gamma ph}(\omega )v_{1ph}(E_{2})\leq 1$, is performed by
using the data of Debbe et al. [18] on \v {C}erenkov ring radii of
electrons, muons, pions and kaons in a RICH detector (see Fig. 4).
The results on this experimental test of the super-coherence
conditions are presented. These S\v {C}R-predictions are verified
experimentally with high accuracy: $\chi ^{2}$/dof =1.47 (see Fig.
5). The scaling law of the ring radii and Yields (see Fig. 5b)
predicted by the S\v {C}R-effect are also experimentally confirmed
with high accuracy.

(iii) The inferred SCR-yield at just \v {C}R-threshold, is of
order of magnitude $Y(SCR)=0.1301$ (see eqn. 19).

(iv)  The influence of medium on the particle propagation
properties is investigated and the refractive properties of
electrons, muons, pions, in the radiator $C_{4}F_{10}Ar$ are
obtained. The refractive indices for this
radiator at $p_{lab}=1GeV$ are as follows: $n_{\mu }=1.001449\pm 0.000098$, $%
n_{\pi }=1.0012593\pm 0.000167$, $n_{K}=1.0214\pm 0.0026$, $n_{p}=1.1066\pm
0.046$. So, we proved that the refractive indices of the particles in medium
are also very important for the RICH detectors, especially at low and
intermediate energies.

(v)  We proved that the anomalous \v {C}erenkov rings observed
recently by Vodopianov et al.[14] at SPS accelerator at CERN
Pb-beam can also be considered as an experimental signature of the
HE-Super-\v {C}erenkov component (see Fig.3).

Finally, we remark that new and accurate experimental measurements of \v
{C}erenkov ring radii, as well as for the anomalous HE-component of S\v {C}R
effect are needed.

\textbf{Acknowledgments.} We would like to thank Prof. Dr. G.
Altarelli for hospitality and fruitful discussions during my stay
in TH-Division CERN-Geneva. This work was supported by CERES
Projects C2-86-2002, C3-13-2003.

\bigskip

\end{document}